\documentclass[11pt]{article}
\usepackage{amsmath}
\usepackage{graphicx}
\usepackage{ulem}
\usepackage{color}
\usepackage{cite}
\begin{document}

\begin{center}
{\Large\bf   Boson-Fermion Duality from Discontinuous Gauge
Symmetry}
\par\vskip20pt
Ji Xu and Shuai Zhao     \\
{\small {\it INPAC, Shanghai Key Laboratory for Particle Physics
and Cosmology, \\
Department of Physics and Astronomy, Shanghai Jiao Tong University,
Shanghai, 200240, China }}
\end{center}
\vskip 1cm

\begin{abstract}

By extending local $U(1)$ gauge symmetry to discontinuous case, we
find that under one special discontinuous $U(1)$ gauge
transformation the symmetric and antisymmetric wave functions can
transform into each other in one dimensional quantum mechanics. The
free spinless fermionic system and bosonic system with $\delta$-type
vector gauge potential are proved to be equivalent. The relation
also holds in higher space-time dimensions.

\vskip 5mm \noindent
\end{abstract}
\vskip 1cm
\par

Can bosons and fermions transform into each other? It is possible in
supersymmetry, a hot candidate for solving the hierarchy problem,
which is still waiting to be examined by high energy experiment.
However, in low dimensional non-relativistic systems, the
equivalence of bosonic and fermionic systems are reported long time
ago\cite{Girardeau:1960,Girardeau:1965,Mattis1965,Coleman:1974bu}.
The specifieness of $1+1$ dimensions leads to bosonization
\cite{Tomonaga:1950zz,Luttinger:1963zz}, which has become a popular
procedure in condensed matter physics. The relation of thermodynamic
properties between $1D$ bosonic and fermionic system is also
analyzed \cite{schmidt:1998,crescimanno:2001}. Massless boson and
fermion theories in $1+1$ dimensional Minkowski and curved
space-time are proved to be equivalent
\cite{freundlich:1972,davies:1978}. Because the spin-statistics
relation is based on relativity, non-relativistic system can escape
from the spin-statistics relation, thus boson can be spinning and
fermion can be spinless. The relation between  boson and spinless
fermion may shed light in the general properties of boson-fermion
duality. In \cite{cheon:pla,Cheon:1998iy} an one-dimensional
fermionic many-body system is found to be equivalent to a system of
bosonic particles interacting through $\delta$-type interaction. The
1+1 dimensional systems with derivative $\delta$-functions and
momentum dependent interactions are also discussed
\cite{BasuMallick:2009km,Grosse:2004rp}, and the roles of symmetry
and supersymmetry played in point interactions have been realized
\cite{Cheon:2000tq,Fulop:1999pf,Ohya:2014lsa,Nagasawa:2002un}.

Thanks to the gauge symmetry, one can tune the phase of a wave
function by gauge transformations. For the difference between boson
and fermion is from the different phase factor when exchanging two
identity particles, it provides the probability of connecting bosons
and fermions from gauge symmetry.

\begin{figure}[!hbt]
\begin{center}
\includegraphics[width=6cm]{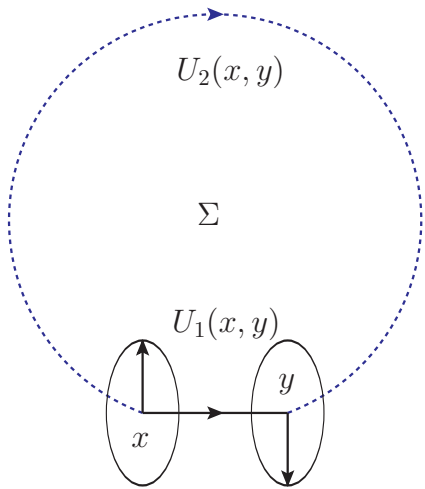}
\end{center}
\caption{Local gauge transformation at two points $x$ and $y$. Two
Wilson line between $x$ and $y$ are presented. The solid line is
``short'' Wilson line while the dashed line denotes the ``longer''
one. When $|x-y|\to 0$, the gauge transformation along $U_1(x,y)$
will be discontinuous. } \label{gauge}
\end{figure}

It is usually considered that the gauge transformation should be
continuous. However, the key idea of the local gauge symmetry is
that the choice of phase at one space-time point should not be
affected by another point. The phase of two points should be
independent to each other even the two points are close enough. Thus
the gauge transformation can be discontinuous. To understand this in
another way, let's consider two points $x$ and $y$ with small
distance. In gauge theory, the gauge transformation of the two
points are related by a Wilson line $U_1(x,y)$, which is shown in
Fig.\ref{gauge}. Generally the difference of the phase $\alpha(x)$
and $\alpha(y)$ is finite. When $x$ and $y$ is close to each other
and keep the difference of the phase $\alpha(x)$ and $\alpha(y)$ as
finite, we will get a discontinuous gauge transformation. One can
also link $x$ and $y$ from another Wilson line $U_2(x,y)$ with
finite length when $|x-y|\to 0$, $U^{\dag}_1$ and $U_2$ then form a
Wilson loop.  When gauge transformation along $U_1(x,y)$ is
discontinuous, the gauge transformation along $U_2(x,y)$ can still
be continuous, which means one can realize discontinuous
transformation through a continuous one.

In the present work we try to connect one-dimensional spinless
fermionic system and bosonic system with discontinuous gauge
transformation. We will show that free fermions can be mapped into
bosons with a $\delta$-function type vector-potential, and vice
versa, which reveal an interesting connection between boson-fermion
duality and gauge symmetry. It is also possible to generalize the
discussions on one-dimensional system to other space-time dimensions
and symmetries.

We start with one-body problem in one dimensional quantum mechanics.
Consider a charged scalar particle in a magnetic field. Its motion
is governed by Schr\"{o}dinger equation with gauge potential
$A_i=(A_0,A_1)$
\begin{eqnarray}
  \left[-\frac{1}{2}\left(\partial_x+ie
  A_1\right)^2+V(x)\right]\phi(x,t)=i\left(\frac{\partial }{\partial
  t}+i eA_0\right)\phi(x,t),\label{scheq}
\end{eqnarray}
For simplicity we adopt the system of natural units and take the
particle mass $m=1$, $\phi(x,t)$ is the wave function, $e$ is the
electric charge of the particle. It is well known that
Eq.(\ref{scheq}) is invariant under $U(1)$ gauge transformation
\begin{align}
  \phi (x,t)\to \phi'(x,t)&=\phi(x,t)e^{i\alpha(x,t)},\\
  A_i(x,t)\to A'_i(x,t)&=A_i(x,t)-\frac{1}{e}\frac{\partial}{\partial
  x_i}\alpha(x,t),
\end{align}
where $i=0,~1$ corresponding to $t,~x$ respectively. As discussed
above, the choice of the phase at one point should not be affected
by another, so it is reasonable that the phase function
$\alpha(x,t)$ can be discontinuous. Here we let $\alpha(x,t)=\pi
\theta(x)$, where $\theta(x)$ is the Heaviside step function,
 $\alpha$ is independent
of $t$. Then we have the discontinuous gauge transformation
\begin{align}
  \phi (x,t)\to \phi'(x,t)&=\phi(x,t)e^{i\pi\theta(x)}\label{gt},\\
  A_0(x,t)\to A'_0(x,t)&=A_0(x,t),\\
  A_1(x,t)\to A'_1(x,t)&=A_1(x,t)-\frac{\pi}{e} \delta(x),
\end{align}
$\delta(x)$ is the Dirac delta function. The transformation in
Eq.(\ref{gt}) can be expressed as
\begin{eqnarray}
  \phi'(x,t)=\left\{
  \begin{aligned}
    -&\phi(x,t),~~~~&x>0;\\
    &\phi(x,t),~~~~&x<0.
  \end{aligned}
  \right.
\end{eqnarray}
Now we assume that $\phi(x,t)$ is an odd function of $x$, i.e.,
$\phi(-x,t)=-\phi(x,t)$, then for $\phi'(x,t)$  we have
$\phi'(-x,t)=\phi'(x,t)$, which means that $\phi'(x,t)$ is an even
function. So we have transform an odd wave function into an even
wave function,  at the cost of introducing a $\delta$-type potential
in the gauge field. Even functions can also be transformed into odd
functions under the same transformation.

\begin{figure}[hbt]
\begin{center}
\includegraphics[width=15cm]{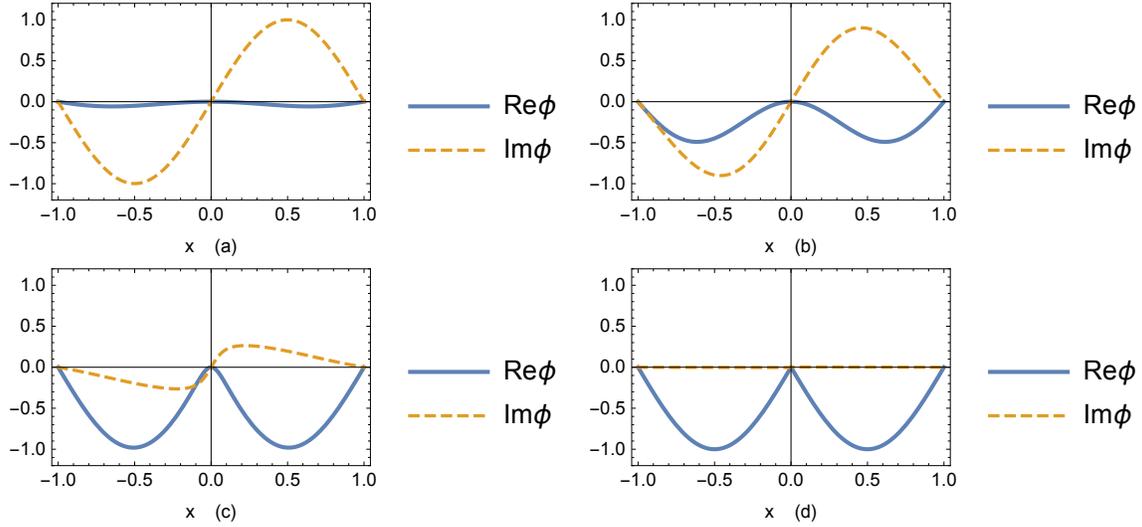}
\end{center}
\caption{The  wave functions of Eq.(\ref{gt}) correspond to $n=2$ at
$\epsilon=10$ (a), $\epsilon=1$ (b), $\epsilon=0.1$ (c) and
$\epsilon=0.001$ (d). The solid lines denote the real parts while
the dashed lines show the imaginary parts of the wave functions.}
\label{wave}
\end{figure}

Before further discussions we should make a few remarks on the
vector potential. In fact there is no magnetic field $B$ in $1+1$
dimensions, even if it exists, charged particle will not couple to
the magnetic field. Further more, it is a little tricky to talk
about ``vector potential'', because $A$ is actually a scalar in
$1+1$ dimensions \cite{elect}. In our case, we just take $A$ as an
auxiliary vector potential, also for the sake of simplicity, we will
set $A_i=0$ and only keep the $\delta$ term in the following
discussions for $1+1$ dimensions. $A$ will make sense in higher
dimensions. With these assumptions, after the gauge transformation,
the new field $\phi'(x,t)$ then satisfies the Schr\"{o}dinger
equation
\begin{eqnarray}
  -\frac{1}{2}\Big[\partial_x-i\pi\delta(x)\Big]^2\phi'(x,t)=i\frac{\partial}{\partial
  t}\phi'(x,t).\label{dteq}
\end{eqnarray}
To see how anti-symmetric wave function transform into symmetric
more clearly, we consider a class of equations which approach to
Eq.(\ref{dteq}) when $\epsilon\to 0$. We restrained the system with
an infinite square potential well, thus one can study the stationary
state problem
\begin{eqnarray}
  -\frac{1}{2}\left[\left(\frac{\partial}{\partial
  x}-i
  \frac{\epsilon}{x^2+\epsilon^2}\right)^2+V(x)\right]\phi'(x)=E_n \phi'(x)
\end{eqnarray}
with
\begin{eqnarray}
  V(x)=\left\{
  \begin{aligned}
    &0,~~~~~~&|x|<a;\\
    &\infty,~~~~~~&|x|>a,
  \end{aligned}
  \right.
\end{eqnarray}
where $E_n$ is the energy level. The eigenfunctions for different
values of $\epsilon$ are shown in Fig.\ref{wave}. When $\epsilon$ is
large, the anti-symmetric imaginary part is dominant. When
$\epsilon$ takes small value the symmetric real part become
important. When $\epsilon\to 0$, $\epsilon/(x^2+\epsilon^2)\to
\pi\delta(x)$, and the anti-symmetric imaginary part of wave
function disappears, we finally arrive at a symmetric wave function.

In \cite{cheon:pla}\cite{Cheon:1998iy}, a fermionic system in an
$\varepsilon(x;c)$ potential is proved to be equivalent with a
bosonic system in $ \delta(x;v)$ potential, with the coupling
strength reversed, i.e., $v=1/c$, here
$\varepsilon(x;c)\equiv\chi(x;-1,0,-1,-4c)$, and $\delta(x;v)\equiv
\chi(x;-1,-v,-1,0)=v \delta(x)$. $\chi$ is defined by
\cite{Cheon:1998iy}
\begin{eqnarray}
  \chi(x;\alpha,\beta,\gamma,\delta)=\displaystyle{\lim_{a\to
  +0}}\left[\left(-\frac{1}{a}+\frac{\gamma-1}{\delta}\right)\delta(x+a)+\frac{1-\alpha\gamma}{\beta a^2}
  \delta(x)+\left(-\frac{1}{a}+\frac{\alpha-1}{\delta}\right)\delta(x-a)\right].
\end{eqnarray}
 In fact what we have discussed above is an extreme case of
such a strong-weak duality. To see this clearly, Eq.(\ref{dteq}) can
be reexpressed as a Schr\"{o}dinger equation with a $\delta^2(x)$
potential
\begin{eqnarray}
  -\frac{1}{2}\frac{\partial^2}{\partial
  x^2}\phi'(x,t)-\pi^2\delta^2(x)\phi'(x,t)=i\frac{\partial }{\partial
  t}\phi'(x,t),
\end{eqnarray}
because $\exp(i\pi\theta(x))=\exp(-i\pi\theta(x))$, the wave
function satisfies the same equation when replacing $\pi$ with
$-\pi$. When $c=a/2\to 0$, $\varepsilon(x;c)\to 0$, Schr\"{o}dinger
equation with $\varepsilon$ interaction then degenerates to a free
particle equation; In another hand, $v=1/c\to \infty$, $v\delta (x)$
then becomes a $\delta$-interaction with infinite coupling strength,
which can be regarded as a $\delta^2(x)$ interaction, because that
when $x$ is close to $0$, $\delta^2(x)\sim \delta(0)\delta(x)$,
which is a $\delta(x)$ interaction with an infinite coupling
strength $\delta(0)$. Thus a fermionic system with zero coupling
strength $\varepsilon$-interaction is equivalent to a bosonic system
with infinite coupling strength $\delta$-interaction.

Now we turn to the two-particle system with a wave function
$\Psi(x_1,x_2,t)$, $x_i(i=1,~2)$ is the coordinate of the $i$-th
particle. Introducing the discontinuous gauge transformation
\begin{eqnarray}
  &&\Psi(x_1, x_2, t)\to \Psi'(x_1, x_2, t)=\Psi(x_1,x_2,t)e^{i\pi\theta(x_1-x_2)},\label{gt2}
\end{eqnarray}
one can immediately get that
\begin{eqnarray}
  \Psi'(x_1, x_2, t)=\left\{\begin{aligned}
     -&\Psi(x_1, x_2, t),~~~~~~&x_1>x_2;\\
     &\Psi(x_1,x_2,t),~~~~~~&x_1<x_2.
  \end{aligned}\right.
\end{eqnarray}
Notice that the sign before the $\delta$ function is opposite for
the two particles. Now we assume that $\Psi(x_1,x_2,t)$ is
anti-symmetric under the exchange of $x_1$ and $x_2$, i.e., a wave
function for fermionic system, then $\Psi'(x_1,x_2,t)$ should not
change when exchanging $x_1$ and $x_2$, which is a wave function of
bosons. Thus we have transformed a wave function of fermion into
bosons under a discontinuous gauge transformation, at the cost of
introducing a $\delta$-type interaction in the derivative of wave
function. The two-body Schr\"{o}dinger equation then reads
\begin{eqnarray}
  -\frac{1}{2}\left[\left(\frac{\partial}{\partial x_1}-i \pi\delta(x_1-x_2)\right)^2+\left(\frac{\partial}{\partial
  x_2}+i\pi\delta(x_1-x_2)\right)^2\right]\Psi'(x_1,x_2,t)=i\frac{\partial}{\partial
  t}\Psi'(x_1,x_2,t).
\end{eqnarray}
Similarly one can consider $n$-particle system, $x_i$ is the
coordinate of the $i$-th particle. Then we perform the gauge
transformation on the wave function
$\Psi(x_1,\cdots,x_i,\cdots,x_n)$:
\begin{eqnarray}
 && \Psi(x_1,\cdots,x_i,\cdots,x_n)\to
  \Psi'(x_1,\cdots,x_n)=\Psi(x_1,\cdots,x_n)\exp{\left[i\pi\displaystyle{\sum_{1\leq i<j \leq n}\theta(x_i-x_j)}\right]}.\label{gtn}
\end{eqnarray}
One can  examine that $\Psi'$ will be multiplied by a $(-1)$ when
exchanging the coordinates of any pair of particles. The wave
function then satisfies the $n-$body Schr\"{o}dinger equation
\begin{eqnarray}
-\frac{1}{2}\displaystyle{\sum_{i=1}^n}\left[\frac{\partial}{\partial
x_i}-i\pi\displaystyle{\sum_{\substack{l,~k\\1\leq l<i<k\leq
n}}}\left(\delta(x_i-x_k)-\delta(x_l-x_i)\right)\right]^2\Psi'(x_1,\cdots,x_n)=i\frac{\partial}{\partial
t}\Psi'(x_1,\cdots,x_n).
\end{eqnarray}
The $\delta$-function in the above equation can be understood as
Pauli exclusion principle, which states that two or more identical
fermions cannot occupy the same quantum state within a quantum
system simultaneously. In one dimensional quantum mechanics, when
the interaction term contains $V=\delta(x_i-x_j)$, particles can not
have the same coordinates at the same time, because that when
$x_i=x_j$, $V$ will be infinite.  In one dimensional system the
quantum states are characterized by the space-time coordinates, so
Pauli exclusion principle states that two fermions can not have the
same coordinates at the same time, which indicates that there is a
$\delta(x_i-x_j)$ interaction in the equation of motion.

The above discussions on one dimensional system can also be
generalized into other space-time dimensions. Now we will show that
for an one-component scalar field with $U(1)$ symmetry, one can also
build up boson-fermion duality in $2+1$ dimensions. Consider a gauge
potential
$\vec{A}=(-\frac{\pi}{2e}\delta(x),-\frac{\pi}{2e}\delta(y))$ and a
symmetric wave function $\phi(\vec{r})$. For any $\vec{r}$ and
$-\vec{r}$, one can link these two points with product of two Wilson
lines $U_{L_2}(\vec{r},0)$ and $U_{L_1}(0,-\vec{r}\,)$:
\begin{align}
  U_{L_2}(\vec{r},0)U_{L_1}(0,-\vec{r}\,)&=\exp\left(ie\int_{L_1} \vec{A}\cdot d
  \vec{l}\right)\exp\left(ie\int_{L_2} \vec{A}\cdot d
  \vec{l}\right)\nonumber\\
  &=\exp\left\{-i\frac{\pi}{2}\int^{0}_{r} ~dR \Big[\cos(\theta+\pi)
  ~\delta\left(R\cos(\theta+\pi)\right)+\sin(\theta+\pi)
  ~\delta\left(R\sin(\theta+\pi)\right)\Big]\right.\nonumber\\
   &~~~~~~~~\,\,\,\,-i\left.\frac{\pi}{2}\int^{r}_{0} ~dR \Big[\cos\theta ~\delta(R\cos\theta)+ \sin\theta ~\delta(R\sin\theta)\Big] \right\}\nonumber\\
  &=-1,
\end{align}
where $L_1$ is a straight line from $-\vec{r}$ to $0$ and $L_2$ is a
straight line from $0$ to $\vec{r}$, $r$ and $\theta$ is the polar
radius and angular respectively. If $\phi(\vec{r})$ is an odd
function, then $\phi'(\vec{r})$ will be an even function. Note that
one can link $\vec{r}$ and $-\vec{r}$ by Wilson lines along other
paths. However, every path links $\vec{r}$ and $-\vec{r}$ will pass
$x$ and $y$ axes once. Every time passing through an axes will
contribute a phase $\exp(i\pi/2)$, finally contributes a factor
$(-1)$ to the relative phase of the two points.

\begin{figure}[!hbt]
\begin{center}
\includegraphics[width=14cm]{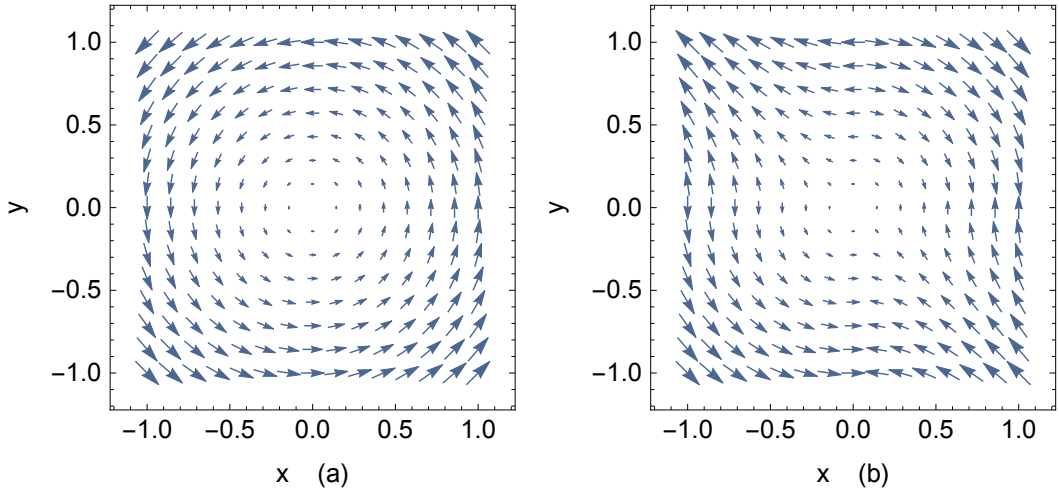}
\end{center}
\caption{The gauge transformation Eq.(\ref{vec}). (a) shows an
anti-symmetric vector wave function in 2 spatial dimensions. (b) is
a symmetric vector wave function which is got by acting
Eq.(\ref{vec}) on (a).} \label{graph3d}
\end{figure}

In 2+1 dimensions we can also consider two components vector wave
function under $U(1)\times U(1)$ symmetry. Consider a wave function
$\Phi(\vec{r},t)=(\phi_1(\vec{r},t),~\phi_2(\vec{r},t))$, where
$\vec{r}=(x,y)$. We perform the gauge transformation
\begin{eqnarray}
  \Phi(\vec{r},t)=(\phi_1(\vec{r},t),~\phi_2(\vec{r},t))&\to &\Phi'(\vec{r},t)=
  (e^{i\pi\theta(x)}\phi_1(\vec{r},t),~~e^{i\pi\theta(y)}\phi_2(\vec{r},t))\nonumber\\
  &=&\left\{\begin{aligned}
    &(-\phi_1(\vec{r},t),-\phi_2(\vec{r},t)),~~~~~~&x>0,~y>0;\\
    &(\phi_1(\vec{r},t),-\phi_2(\vec{r},t)),~~~~~~&x<0,~y>0;\\
    &(\phi_1(\vec{r},t),\phi_2(\vec{r},t)),~~~~~~&x<0,~y<0;\\
    &(-\phi_1(\vec{r},t),\phi_2(\vec{r},t)),~~~~~~&x>0,~y<0.
  \end{aligned}\right.\label{vec}
\end{eqnarray}
If $\Phi(\vec{r},t)$ is an odd function with $\vec{r}$, i.e.
$\Phi(-\vec{r},t)=\Phi(\vec{r},t)$, then
$\Phi'(-\vec{r},t)=-\Phi'(\vec{r},t)$, which means that
$\Phi'(\vec{r},t)$ is an even function. Fig.\ref{graph3d} shows that
an odd vector wave function transforms into an even vector wave
function under Eq.(\ref{vec}).
The gauge transformation on gauge potential $A_i$ then becomes
\begin{align}
    {A}_0(\vec{r},t)&\to A'_0(\vec{r},t),\\
    {A}_x(\vec{r},t)&\to {A}'_x(\vec{r},t)={A}_x(\vec{r},t)-\frac{\pi}{e}\delta(x),\\
    {A}_y(\vec{r},t)&\to {A}'_y(\vec{r},t)={A}_y(\vec{r},t)-\frac{\pi}{e}\delta(y).
\end{align}
One can also construct many-body equations in 2+1 dimensions, just
as what we have done for $1+1$ dimensional case. These discussions
can also be generalized to higher space-time dimensions.

The Aharonov-Bohm effect \cite{Aharonov:1959fk} states that outside
the region where magnetic field $B$ is confined, the vector
potential may still causes observable effects. The above discussions
also reveal some novel properties of vector potential. If the vector
potential is confined in a $n-1$ dimensional hyperplane in $n$
spatial dimensions, it can also have physical effects. For certain
configurations they may affect the shape of wave functions. In
addition, this work only consider scalar wave functions governed by
Schr\"{o}dinger equation. Similar discussions can be performed to
Dirac equations as well. Further more, in this work we only consider
$U(1)$ and $U(1)\times U(1)$ groups.  It should be also interesting
of generalizing this work to non-Abelian symmetries. We leave these
for a further research.

To summarize, we build up boson-fermion duality under the spirit of
local gauge symmetry. The non-relativistic bosonic (fermionic)
system can be mapped into a fermionic (bosonic) system with
$\delta$-type gauge interactions. This duality may open a door to a
deeper understanding of the relationship between exchange statistics
and gauge symmetry.

\par\vskip20pt

\noindent
{\bf Acknowledgments}
\par
We thank Dr. Jian-Ping Dai for valuable discussions.

\par\vskip40pt

\par

\end{document}